\documentclass[fleqn]{annalen}
\usepackage{graphics}
\usepackage{latexsym}
\begin{document}
\newcommand{\volume}{xx}              
\newcommand{\xyear}{1999}            
\newcommand{\issue}{xx}             
\newcommand{\recdate}{dd.mm.yyyy}   
\newcommand{\revdate}{dd.mm.yyyy}    
\newcommand{\revnum}{0}              
\newcommand{\accdate}{dd.mm.yyyy}  
\newcommand{\coeditor}{xx}           
\newcommand{\firstpage}{1}           
\newcommand{\lastpage}{4}            
\setcounter{page}{\firstpage}        
\newcommand{\keywords}{disordered solids, metal-insulator transitions,
thermoelectric conduction} 
\newcommand{\PACS}{61.43.--j, 71.30.+h, 72.15.Cz}
\newcommand{\shorttitle}{C. Villagonzalo et al., Transport Properties near the
Anderson transition} 
\title{Transport properties near the Anderson transition}
\author{C. Villagonzalo, R.\ A.\ R\"{o}mer, and M. Schreiber} 
\newcommand{\address}
  {Institut f\"{u}r Physik, Technische Universit\"{a}t, D-09107 Chemnitz, 
  Germany}
\newcommand{\email}{\tt villagonzalo@physik.tu-chemnitz.de} 
\maketitle
\begin{abstract}
  The electronic transport properties in the presence of a temperature
  gradient $\nabla T$ in disordered systems near the metal-insulator
  transition (MIT) are considered.  The d.c. conductivity $\sigma$,
  the thermoelectric power $S$, the thermal conductivity $K$ and the
  Lorenz number $L_0$ are calculated for the three-dimensional (3D)
  Anderson model of localization using the
  Chester-Thellung-Kubo-Greenwood formulation of linear response. We
  show that $\sigma$, $S$, $K$ and $L_0$ can be scaled to
  one-parameter scaling curves with a single scaling parameter $k_B
  T/|(\mu-E_c)/E_c|$.
\end{abstract}

\section{Introduction}

In this paper we shall demonstrate that in the Anderson model of
localization near the metal-insulator transition (MIT) the d.c.\ 
conductivity $\sigma$, the thermoelectric power $S$, the thermal
conductivity $K$ and the Lorenz number $L_0$ obey scaling.  

A fundamental model often used in dealing with transport phenomena
of disordered media is given by the Anderson Hamiltonian \cite{anderson}.
Investigations of this model yield that the energy $E$ dependent
conductivity $\sigma(E)$ behaves as
\begin{equation}
\sigma(E)=\left\{ \begin{array}{cc}
\sigma_{0}\left|1-\frac{E}{E_{c}}\right|^{\nu}, & \quad |E| < E_{c}, \\
0,                                   & \quad |E|\geq E_{c},
\end{array}\right. \label{dc_cond}
\end{equation}
in energy regions near the mobility edge $E_c$ at which the MIT occurs
\cite{kramer}. Here $\sigma_0$ is a constant and $\nu$ is a universal
critical exponent. Using Eq.\ (\ref{dc_cond}) we have been able to
study the temperature $T$ dependence of $\sigma$, $S$, $K$ and $L_0$
near the MIT \cite{villa} using linear response theory \cite{chester}.
Here we present our observations of  their scaling features.

\section{Calculating the transport properties}

In the presence of a small temperature gradient $\nabla T$ in an open
circuit, $S$ is the coefficient of proportionality between $\nabla T$
and the electric field it induces. The coefficient $K$ relates $\nabla T$
to the heat current while the Lorenz number $L_0$ measures
the ratio between $K$ and the product $\sigma T$. The transport
properties are defined in the framework of linear response theory as
\begin{eqnarray}
\sigma = L_{11}, \hspace*{2.2cm}  & &
S = \frac{L_{12}}{|e|TL_{11}}\;,  \nonumber\\[-.5ex] \label{lreq}\\[-0.3cm]
K=\frac{L_{22}L_{11}-L_{21}L_{12}}{e^{2}TL_{11}}\,, & \mbox{and} &
L_{0}=\frac{L_{22}L_{11}-L_{21}L_{12}}{(k_{B}TL_{11})^2}\,, \nonumber 
\end{eqnarray} 
where $e$ is the electron charge and $k_B$ is Boltzmann's
constant.  For a noninteracting system with no inelastic scattering
processes present, such as phonon mediated hopping, the kinetic
coefficients $L_{ij}$ are given in the Chester-Thellung-Kubo-Greenwood
formulation \cite{chester} as
\begin{equation} 
L_{ij}= (-1)^{i+j}\int_{-\infty}^{\infty} A(E)
\left[E-\mu(T)\right]^{i+j-2} \left[-
\frac{\partial f(E,\mu,T)}{\partial E} \right] dE\;, 
\label{coeff}
\end{equation} 
where $i,j=1,2$, $\mu$ is the chemical potential of the system,
$f(E,\mu,T)$ is the Fermi distribution function, and $A(E)$ describes the
system dependent features. In the Anderson model, one finds $A(E)$ to be
equal to the critical behavior (\ref{dc_cond}) of $\sigma$ near the MIT
\cite{imry}.

Previously, we have numerically determined $\mu(T)$ for the 3D
Anderson model from its density of states \cite{villa}. Using that
result, i.e., $\mu(T)\propto T^2$, it is straightforward to determine
the kinetic coefficients (\ref{coeff}) and, consequently, the transport
properties (\ref{lreq}).

In order to compare our results with experiments we have
chosen the energy units to be in electron volts. For the same reason,
we measure $\sigma$ in units of $\Omega ^{-1}\mbox{cm}^{-1}$.  
In this paper we only show the results obtained for disorder $W=12$
where W is the width of the box distribution of randomly chosen
potential energies in the Anderson model of localization. We have
obtained similar results for other disorders not too close to the critical
disorder $W_c\approx 16.5$ \cite{schreiber,slevin} and where there are
no large fluctuations in the density of states.
For $W=12$, the corresponding mobility edge is at $E_c=7.5$ \cite{schreiber}.
The value of $\nu$ is set to 1.3 in agreement with numerical
results \cite{hofstetter}. Note that this choice of $\nu$ determines
the magnitude of the transport properties but does not change
their $T$ dependence \cite{villa} upon which their scaling is based.

\section{Scaling Features}

Wegner \cite{wegner} was the first to show for disordered
noninteracting electron systems that $\sigma$ can be
written close to the MIT in a scaling form as
\begin{equation}
\sigma(t,\Omega)=b^{2-d}F(b^{1/\nu}t,b^{z}\Omega)\;.
\label{sigscale1}
\end{equation}
Here $F$ is a system dependent function of the dimensionless parameter
$t$ giving the distance from the critical point, $\Omega$ is an
external parameter such as the frequency, $b$ is a scaling parameter,
$d$ is the dimension and $z$ is the dynamical exponent. For the
noninteracting case, $z=d$ \cite{belitz}.  In this paper, the
appropriate parameters are $d=3$, $t = |(\mu-E_c)/E_c|$, $\Omega = T$
and $b=t^{-\nu}$ and Eq.\ (\ref{sigscale1}) simplifies to
\begin{equation}
\frac{\sigma(t,T)}{t^{\nu}}= F\left(\frac{T}{t^{\nu z}}\right).
\label{sigscale2}
\end{equation}

\begin{figure}[t]
\hspace{1mm}
\centerline{
\resizebox{5.6cm}{!}{\includegraphics{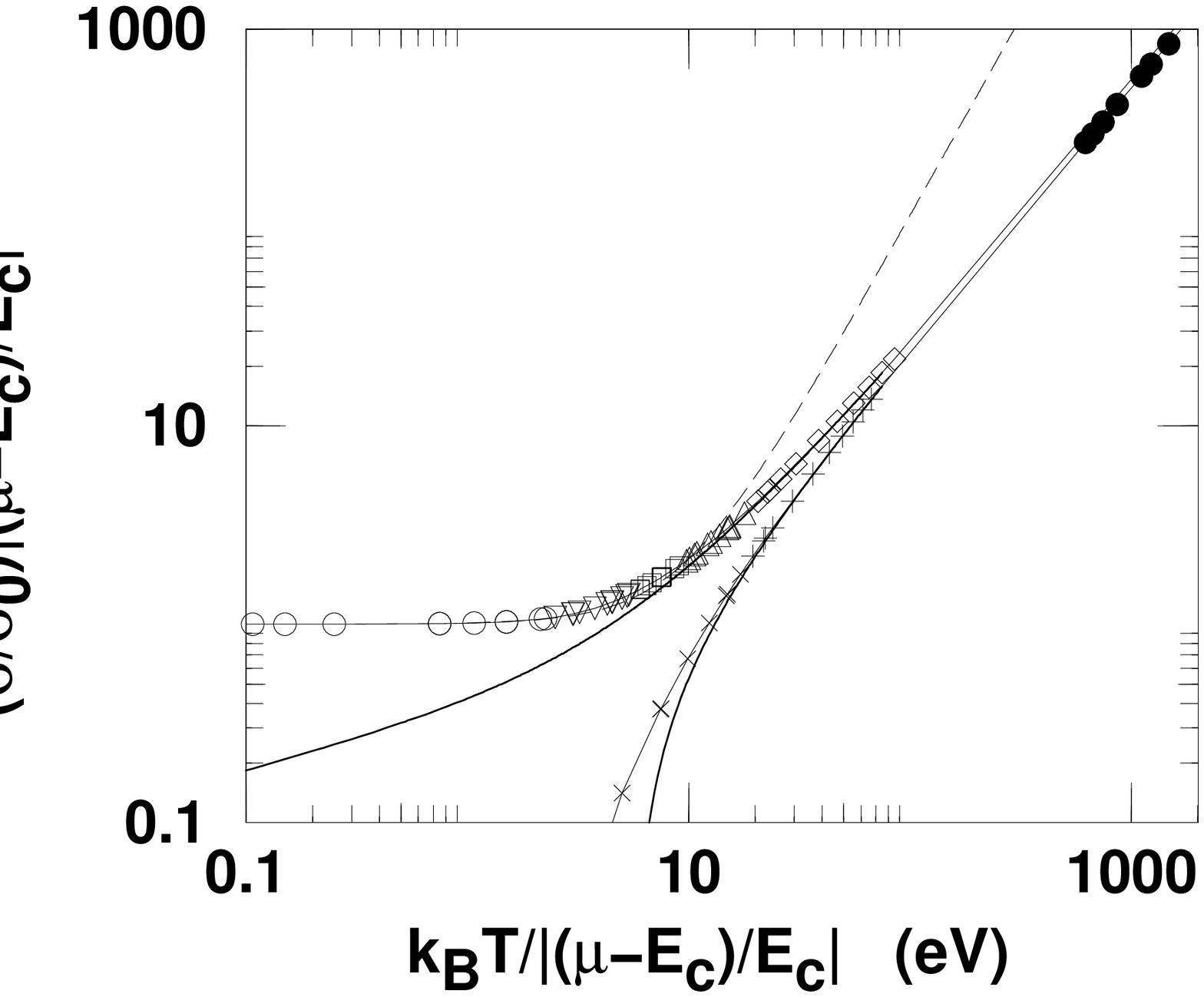}}
\hspace*{0.5cm} 
\resizebox{5.6cm}{!}{\includegraphics{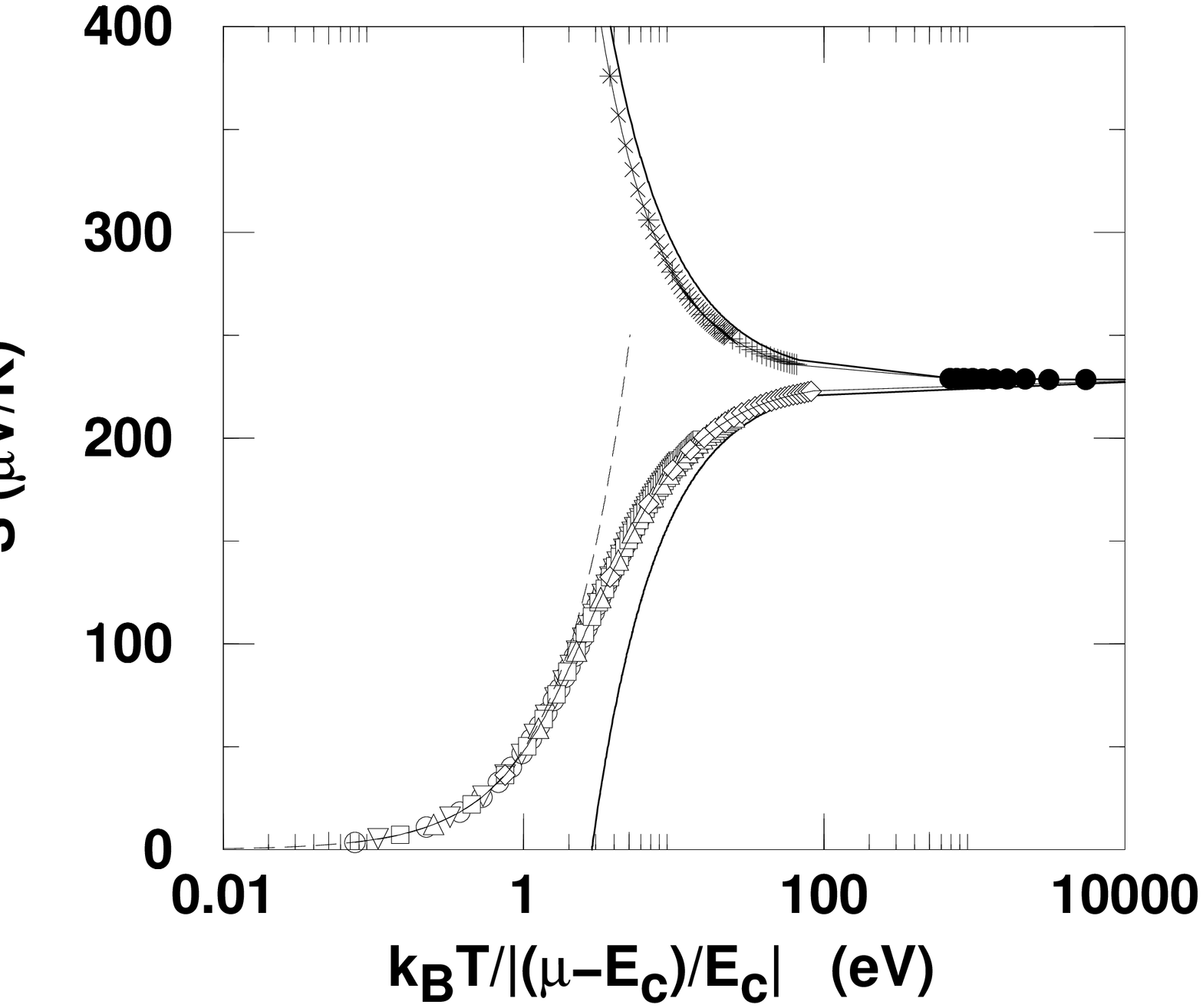}} 
}
\caption{Left: The d.c.\ conductivity obtained in the metallic regime:
  $E_F-E_c=-0.01$ eV ($\circ$), $-0.007$ eV ($\bigtriangledown$),
  $-0.005$ eV ($\Box$), $-0.003$ eV ($\bigtriangleup$), $-0.001$ eV
  ($\Diamond$), in the critical regime: $E_F-E_c=0$ eV ($\bullet$),
  and in the insulating regime: $E_F-E_c=0.001$ eV ($+$), $0.003$ eV
  ($\times$). The thin solid line is the scaling curve onto which all
  data from the different energy regimes collapses.  The thick solid
  line denotes a high temperature expansion \protect\cite{villa,imry}
  while the dashed line is the Sommerfeld expansion
  \protect\cite{imry,castellani} for $E_F-E_c=-0.01$ eV.  Right: The
  scaling plot of the thermoelectric power obtained from
  the different regions in the same way as for $\sigma$. }
 \label{fig1}
\end{figure}

In Fig.\ \ref{fig1} we show that the $\sigma$ data from the metallic
($|E_F| < E_c$), the insulating ($|E_F| > E_c$) and the critical ($E_F
= E_c$) energy regions collapse onto a single scaling curve
when plotted as a function of $k_BT/|(\mu-E_c)/E_c|$.
In the right hand side of Fig.\ \ref{fig1}, we illustrate similar scaling
of $S$. Note that there is no pre-factor in the scaling form such as
in $\sigma$ since $S$ becomes independent of $T$ at the MIT as $T
\rightarrow 0$.  It was pointed out in Ref.\ \cite{imry} that $S$ for
high and low $T$ might scale in terms of $k_B T/|E_F-E_c|$.  This is
approximately valid \cite{villa} although the proper scaling variable
should be $k_B T/|(\mu-E_c)/E_c|$.  The scaling curve for $S$
converges at the value $228.4\,\mu$V/K as predicted in Ref.\ 
\cite{enderby} for $\nu=1.3$.  In the metallic regime, our result
agrees with the Sommerfeld expansion result at low $T$. We find that
$S$ grows infinitely large in the insulating regime. This can be
attributed to the decrease in charge carriers which contribute to
a current.

\begin{figure}[t]
\hspace*{2mm}
\centerline{
\resizebox{5.7cm}{!}{\includegraphics{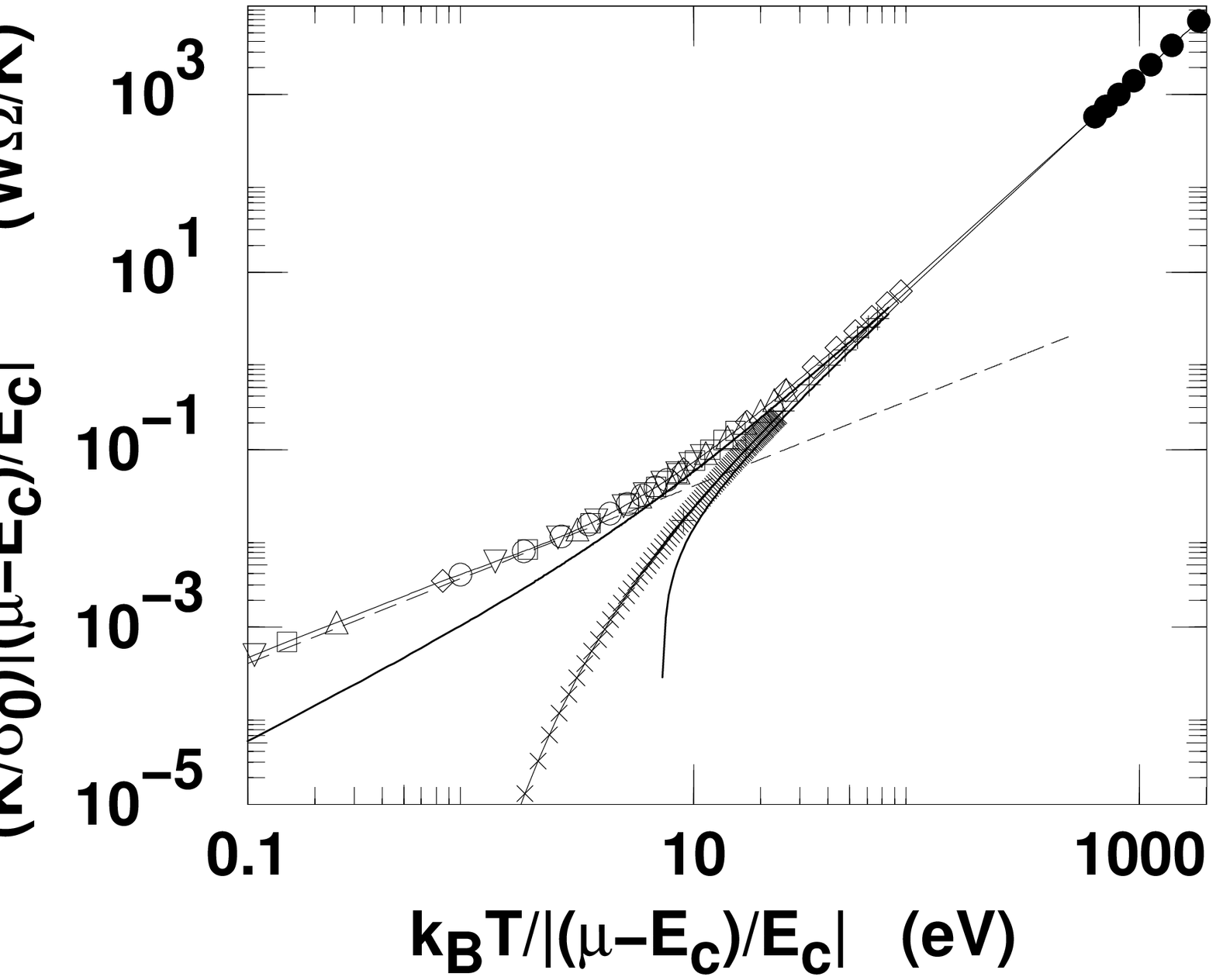}}
\hspace*{0.5cm}
\resizebox{5.7cm}{!}{\includegraphics{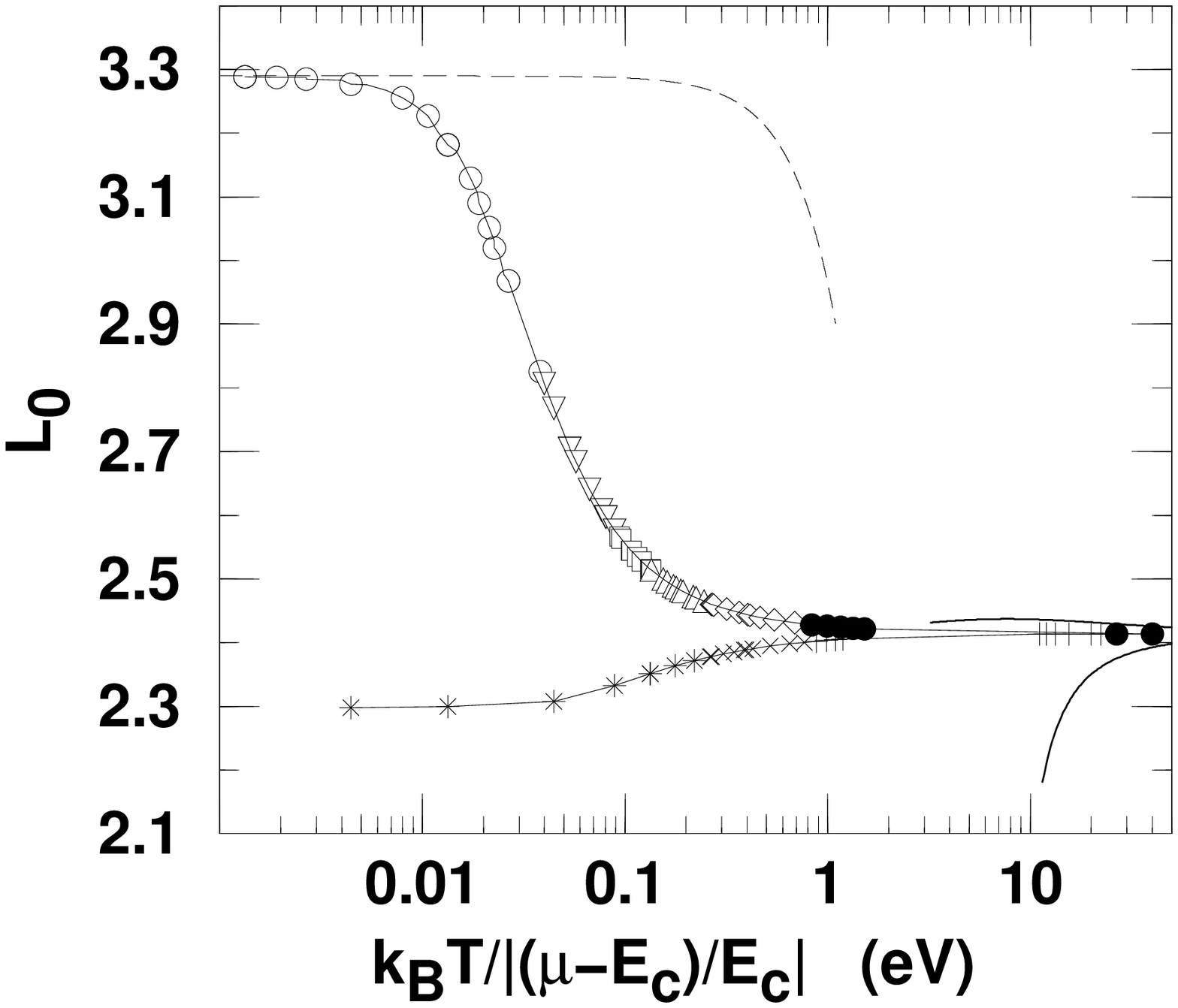}} 
}
\caption{Left: Scaling plot of the thermal conductivity obtained from the same 
  energy regions as given in Fig.\ \protect\ref{fig1}.  The thin solid
  line is the scaling curve onto which all data from the different
  energy regimes collapses.  The thick solid line denotes a high
  temperature expansion \protect\cite{villa,imry} while the dashed
  line is the Sommerfeld expansion \protect\cite{imry,castellani} for
  $E_F-E_c=-0.01$ eV.  Right: The scaling of the  Lorenz
  number.  The additional symbol ($*$) denotes data
  obtained deeper in the insulating regime:
  $E_F-E_c=0.01$ eV.}
 \label{fig2}
\end{figure}

The corresponding scaling curves for $K$ and $L_0$ are shown in Fig.
\ref{fig2}.  They also scale with respect to $k_B T/|(\mu-E_c)/E_c|$.
The normalization factor in $K$ is due to the fact that
$K\propto k_BT^{\nu+1}$ as $T\rightarrow0$ at the critical regime. 
$L_0$ does not require a pre-factor since it is
independent of $T$ at the MIT \cite{enderby}.
The result for $L_0$ shows that it approaches $\pi^2/3$ in the metallic
regime as expected in the Sommerfeld free electron theory of metals.
In the critical regime, $L_0$ converges to $2.414$ which is the value
for $\nu=1.3$ predicted in Ref.\ \cite{enderby}. Due to the
exponential decay of the derivative of the Fermi function as $T
\rightarrow 0$, $L_0$ approaches $\nu+1=2.3$ in the insulating regime.


To summarize, we achieve scaling for the four transport quantities
$\sigma$, $S$, $K$ and $L_0$ with respect to a single parameter $k_B
T/|(\mu-E_c)/E_c|$.  This factor stems from the form of $L_{ij}$ and
from the fact that the explicit $T$ dependence enters via $\mu$ and
the Fermi distribution function.  Comparing our results from $\sigma$
in Fig.\ \ref{fig1} with Eq.\ (\ref{sigscale2}) we obtain $\nu z = 1$.
With $\nu = 1.3$ this yields a dynamical exponent $z=0.77$ which is to
be compared to $z=d=3$ as expected from scaling arguments
\cite{wegner,belitz}.  Moreover, our result that $\nu z =1$ seems to
violate the Harris criterion \cite{harris} which requires $\nu z > 2$.
However, this observation needs to be examined in more detail since
the Harris criterion might be altered for our choice of $T$ dependence
via the Fermi distribution.

\vspace*{0.25cm} \baselineskip=10pt{\small \noindent We thank D.\ 
  Belitz, T.\ R.\ Kirkpatrick, A. MacKinnon and T. Vojta for useful
  discussions. Financial support by the DFG through
  Sonderforschungsbereich 393 is gratefully acknowledged.}

\end{document}